# Mode transition control of large-size tiltrotor aircraft

**George Kirst, Xinhua Wang**
Aerospace Engineering,
University of Nottingham, UK
Email: wangxinhua04@gmail.com

**ABSTRACT**
Tiltrotors are an aircraft concept with the ability to rotate their rotors freely, achieving vertical take-off and fast forward flight. The combination of helicopter and fixed-wing flight into one aircraft provides versatility in mission selection, yet challenges persist in their construction and control. Tiltrotor aircraft can operate in three primary modes: helicopter, fixed-wing, and transition, with the transition mode facilitating the shift between helicopter and fixed-wing flight. However, control within this transition region is inherently challenging due to its non-linear nature, hence tiltrotors have been predominantly limited to military applications. Thus, this paper aims to explore transition mode control for a large-size tiltrotor aircraft, tailored to civil applications.
A novel, large-sized, tiltrotor concept is presented, accompanied by a derived mathematical model describing the aircrafts behaviours. A PID control method has been used to control the height, pitch, and velocity variations within the transition mode with secondary control loop developed to control the tilt angle during transition. The derived model and control are then implemented within a MATLAB simulation, where the control method was iterated to improve performance. The results show a full transition was achieved in under 14 seconds, where altitude variations were kept below 10 metres. Though the transition mode control was successful, a collective look at the data showcases issues with assumptions as well as thrust discontinuities. The implications of these results are discussed, with suggested improvements proposed for future work.

## 1. INTRODUCTION

This paper considers the transition control for a large-sized tiltrotor aircraft. For this problem, an autonomous control solution is proposed for a novel, tiltrotor aircraft configuration. Through the methods proposed, this paper aims to produce a robust control solution that can stabilize the aircraft through transition, with the goal to improve the adoption of tiltrotor aircraft in wider industry.

### 1.1 Tiltrotor Aircraft

Sustained flight in civil and military aviation is dominated by helicopters and fixed-wing aircraft, with more novel concepts often being overlooked due to technological limitations [1]. Several underlying restrictions of these aircraft, has pushed attention to more novel configurations, with the aim of removing these limitations on future air travel.

Tilt Rotors in the same way are an aircraft concept with the primary ability to direct their rotors at varying angles to achieve both forward and vertical thrust. This allows the benefits of vertical take-off and landing (VTOL) with fast forward flight, [2, 3, 4] effectively combining fixed-wing and helicopter concepts into one aircraft. Given their benefits, the concept has attracted attention from the military sector, notably the Osprey-V22 and the AW609. Yet despite the benefits of the concept, the high development costs due to several challenges in design and control [1, 5], have resulted in lacking success in the civil aviation sector.

### 1.2 Transition

Unique to tiltrotor aircraft is the transition mode, wherein the aircraft shifts between helicopter and fixed-wing flight configurations. During this phase, the rotor angle changes from 90-0 degrees, which vectors the thrust forward, accelerating the aircraft and increasing its forward flight velocity.

While helicopter and fixed-wing flight modes are widely recognized, the transition mode presents a unique set of challenges with limited examples of successful implementations. The complexity and non-trivial nature of transitioning from helicopter to fixed-wing mode underscores the importance of accurately capturing the behaviour's exhibited within this mode.

### 1.3 Challenge of Control

Transitioning between helicopter and fixed-wing flight modes poses a significant challenge for control, primarily due to the highly coupled flight dynamics [3, 4] and nonlinearities inherent with the transition phase.

As the aircraft changes from vertical to horizontal flight, altitude and pitch dynamics exhibit rapid change. Notably, the transition involves replacing the lift generated with the rotors, by the lift generated by the fixed-wing, leading to significant variation in the systems aerodynamic characteristics [3]. The control system must adequately manage these dynamic shifts, particularly as thrust levels fluctuate prominently throughout the transition process.

To complicate the issue further, the large propellers, required for lift during helicopter mode, induces complex flow patterns around the wing [6]. These flow dynamics, which contribute to system instabilities, are then exacerbated by external disturbances like wind gusts, thereby diminishing overall stability.

The inherently non-linear nature of this control problem highlights the imperative for a comprehensive understanding of the aerodynamics and the dynamic behaviour throughout transition. Addressing this control challenge therefore requires sophisticated control strategies to implement a successful control method.

## 1.4 PID & Gain Scheduling

While (proportional, integral, differential) PID control methods are widely used, they struggle to adapt to the changing aircraft dynamics during mode transition. With varied flight conditions, a single set of controller gains lacks consistent responsiveness across the entire flight envelope [2, 7]. Scheduling the control gains based on the current flight condition (Gain Scheduling) can be used to tune the responsiveness at each flight conditions, thus providing stable and responsiveness control throughout the flight envelope.

## 1.5 Conversion Corridor

To ascertain aircraft stability, a conversion corridor can be established. This corridor describes a range of forward flight velocities and tilt angles within which the aircraft can maintain stability during transitioning.

Adhering to this corridor is critical, as deviating from it implies an inability to sustain stability [8]. By constructing a conversion corridor outlined in section 5, we effectively define the stability region, as the aircraft tilts [6], enabling the determination of a stable desired tilt angle for every forward flight velocity. However, this only describes the points at which stability is feasible, not points of stability, hence this tool must be used in conjunction with control methods such as gain scheduling, to build a suitable controller.

## 1.6 Large-Size

For small sized tiltrotor aircraft, the act of transition control is simple [9]. Due to their small size, the aircraft can produce enough thrust to both counteract weight, as well as accelerate forward. However, for large aircraft the maximum thrust to weight ratio becomes a significant factor, as an increase in thrust necessitates an increase in engine and propeller size, which negatively impacts the efficiency of forward flight.

Thus, given our project aims we have selected a large-scale aircraft, such that control becomes non-trivial.

## 1.7 Aims & Objectives

Our project aims to develop a robust control method, facilitating stable transition between vertical and horizontal flight modes. This endeavour seeks to deepen our understanding of the transition mode, enhancing the applicability of tiltrotor technology in future aircraft designs.

Objectives include designing a technically feasible aircraft, developing a complete mathematical model to describe dynamic behaviour, implementing a control method, and conducting simulations for performance evaluation. Through completion of these objectives, we will have assessed the viability of our derived control method, and thus satisfied our overall project aims.

## 2. AIRCRAFT CONFIGURATION

The following section highlights the selected aircraft, for which control will be implemented. It is important to consider the aircrafts feasibility, as our conclusions on stability will be negatively impacted.

## 2.1 General Configuration

A large 6 engine, 12 tonne cargo aircraft has been developed with an outline of the model shown in Figure 1. Four engines are located on the aircraft wing whilst two are located on the tail, providing a significant lever arm to balance the aircraft.

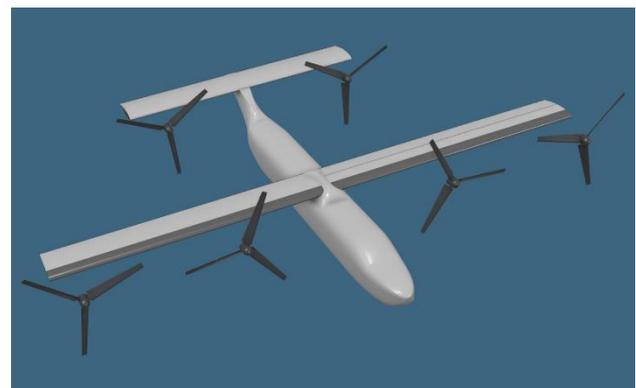

**Figure 1 Chosen Aircraft, Layout of Key Components**

The configuration was chosen from among several designs based on several performance criteria, notably a trade-off between performance and design feasibility. Increasing the number of rotors was seen to benefit the stability and redundancy of the aircraft. However, it was impractical in terms of design implementation, thus a 6-engine design was selected as a compromise between them.

## 2.2 Initial Sizing

The aircrafts maximum take-off weight (MTOW) was based on a defined payload of 2 metric tonnes and a range based on historic aircraft data. This MTOW was then used to approximate the maximum required thrust for each engine based on a safety factor of 1.5. The design was then iterated to achieve a reasonable aircraft configuration. Methods to size the aircraft were taken from Raymer [10] with a summary of aircraft parameters displayed in Table 1.

**Table 1: Key Aircraft Parameters**

| Parameter | Value |
|---:|---|
| MTOW | 12,000kg |
| Wingspan | 26m |
| Rotor Diameter | 5.4m |
| Design Range | 500Nm |
| Design Altitude | 25,000ft |
| Disk loading | 130kgm$^{-2}$ |
| Cruise Velocity | 125ms$^{-1}$ |

## 2.3 Wing Sizing

Aircraft data from similar-sized fixed-wing aircraft were utilized to estimate a wing planform. Subsequently, data from the MS317-IL aerofoil, obtained from an aircraft with a similar payload size, was employed to calculate the required wing area based on cruise requirements. Given the relatively low designed flight velocity of the aircraft, incorporating a taper into the wing design would not significantly enhance aerodynamic performance. Therefore, a square wing with an aspect ratio of 12 was selected to finalize the wing planform.

## 2.4 Rotor & Engine Sizing

The size of the rotors was determined by comparison of a disk loading, taken from historic data, and a blade element method [11], developed using a flat plate aerofoil assumption. These two methods were then used to derive the power and thrust requirements for each engine. For these requirements, the engine (PT6C-67A) was selected and used to estimate inertial data for the aircraft mathematical model. This engine was previously used in the AW609 tiltrotor aircraft, and such is known to be suitable for tiltrotor application.

Vortex ring generation [12] was assumed to be negligible, with the spacing of the rotors increased to mitigate their effect and ensure the validity of this assumption. However, to fully confirm this assumption, further analysis, through physical and numerical testing would be necessary, outside the scope of this paper.

## 2.5 Wing & Tail Placement

The wing and tail were placed along the fuselage based on the two stability points for helicopter and fixed-wing modes, where the sum of forces and moments equal zero. Helicopter mode constrained the tail mounted engines to be 2 times the distance from the CG position than the wing mounted engines due to the wing having more thrust capability. Whereas the fixed-wing mode constrained the relative heights of each engine about the CG position in the vertical direction in addition to ensuring that a trim position for the aircraft was attainable.

## 2.6 Sustainability

As the configuration is focused on civil application, several design decisions have been made with implications for current and potential future sustainability initiatives within the tiltrotor aircraft domain. For instance, the decision to incorporate a greater number of engines than typical tiltrotors aircraft was to facilitate electrification. Increasing the number of engines decreases their individual power requirements, aligning with the trend towards electrification in aviation. Tiltrotors, with their VTOL capability and low infrastructure requirements, are well-suited to urban air environments [13], where electrification has become a focal point in sustainability efforts.

Despite the emphasis on control in this paper, the broader context of sustainability in aircraft design remains pertinent. As the industry moves towards developing more sustainable aircraft, the challenges of control persist. Therefore, design decisions made in this paper are intended to ensure that its focus remains relevant for future aircraft designs, even as sustainability considerations become increasingly prominent in the field.

## 3. MATHEMATICAL MODELLING

As in similar works [6, 4, 14], to ensure the validity of the control system a robust mathematical model was developed. The resolution of the model affects the validity of the control system and thus we must ensure the model considers all parameters within the scope of the chosen configuration.

## 3.1 Degrees of Freedom (DOF)

To limit the scope of the problem we will only consider a 3 DOF problem, constrained within the X-Z plane. The force component acting on the aircraft in the X and Z directions and the pitching moment acting about the aircraft centre of gravity (CG) are defined as $F_X$, $F_Z$ and $M_\theta$ respectively.

## 3.2 Wing & Tail

Assuming that a control method is capable of stabilising altitude, allows us to approximate the flightpath angle to be zero. Using this assumption, the forces and moments from the wing and tail $F_{LX}$, $F_{LZ}$ and $M_{L\theta}$ are formulated in Eq. (1). The expression is based on the lift generated by the wing and tail ($L_w$, $L_t$) as well as the geometry defined by the configuration design where $X_w$ and $X_t$ are the distances from the centre of pressure of the wing and tail to the CG position.

$$\begin{bmatrix} F_{LX} \\ F_{LZ} \\ M_{L\theta} \end{bmatrix} = \begin{bmatrix} -D \\ -(L_w + L_t) \\ L_w X_w - L_t X_t + DY_D \end{bmatrix} \quad (1)$$

Similarly, the aircraft Drag force $D$ and distance from the CG to the centre of drag $Y_D$ is included to capture the aerodynamic forces acting on the wing and tail.

## 3.3 Thrust

We can group the four wing mounted engines and the two rear together and define them as $F_1$ and $F_2$ respectively, where the max thrust available for $F_1$ is two times greater than $F_2$. The force contribution from the engines are captured in Eq. (2), with pitch and tilt angle represented by $\theta$ and $\tau$, respectively.

$$\begin{bmatrix} F_{TX} \\ F_{TZ} \\ M_{T\theta} \end{bmatrix} = \begin{bmatrix} (F_1 + F_2)\cos(\theta + \tau) \\ -(F_1 + F_2)\sin(\theta + \tau) \\ F_1 g(\tau) - F_2 h(\tau) \end{bmatrix} \quad (2)$$

The lever arm of each thrust group, are described by linear functions $g(\tau)$ and $h(\tau)$ based on the constant, configuration geometry, and the time-varying, tilt angle.

## 3.4 Rotor Inertia

As each engine rotates by an angle $\tau$, the rotational acceleration, $\ddot{\tau}$, induces an adverse torque on the body of the aircraft.
The scale of the moment produced is proportional to the rotational acceleration and the inertia of the engine, $J_r$, attained in section 2.4 and shown in Eq. (3).

$$\begin{bmatrix} F_{\tau X} \\ F_{\tau Z} \\ M_{\tau \theta} \end{bmatrix} = \begin{bmatrix} 0 \\ 0 \\ J_r \ddot{\tau} \end{bmatrix} \quad (3)$$

## 3.5 Disturbance

The model has outlined the aircraft's anticipated behaviour under normal conditions. However, to accommodate for known variables, such as wind gusts, we need to enhance the model. Introducing a disturbance term, $\delta$, into each aspect of the model, allows us to accurately predict the aircraft's response to these environmental factors. This refinement ensures a more thorough representation of the aircraft's dynamics in real-world scenarios.

## 3.6 Summation of Body Forces

Combining the forces derived in sections 3.2-3.5, we get Eq. (4), with $W$ being the weight of the aircraft acting downward.

$$\begin{bmatrix} F_X \\ F_Z \\ T_\theta \end{bmatrix} = \begin{bmatrix} F_{LX} \\ F_{LZ} \\ M_{L\theta} \end{bmatrix} + \begin{bmatrix} F_{TX} \\ F_{TZ} \\ M_{T\theta} \end{bmatrix} + \begin{bmatrix} 0 \\ 0 \\ M_{\tau\theta} \end{bmatrix} + \begin{bmatrix} 0 \\ W \\ 0 \end{bmatrix} + \begin{bmatrix} \delta_X \\ \delta_Z \\ \delta_\theta \end{bmatrix} \quad (4)$$

These equations, govern the aircrafts motion in altitude, pitch and longitudinal position for all three flight modes.

## 4. CONVERSION CORRIDOR

Establishing a conversion corridor, defines the bounds within which a stable transition can be achieved [6]. The following procedure outlines the formulation of each constraint, culminating in the development of a conversion corridor tailored to our prescribed aircraft design.

### 4.1 Normal Operative Modes

The positions of helicopter and fixed-wing flight modes are highlighted in Figure 2, wherein the transition mode is described as any angle of tilt, bounded by these two flight modes.

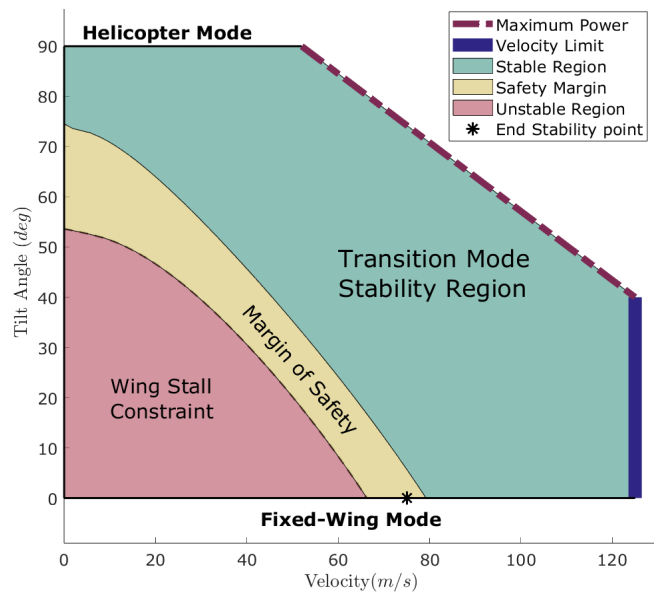

**Figure 2 Aircraft Conversion Corridor**

While the velocity in helicopter mode can be negative, this has not been represented, as conversion can only occur at positive velocities.

## 4.2 Wing Stall Constraint
In figure 2, an unstable region is shown, characterised by a wing stall constraint, where the conditions described in Eq. (5), are not satisfied.

$$F_{TZ(MAX)} + F_{LZ} \geq W \quad (5)$$

Here, the maximum component of thrust acting upward, $F_{TZ(MAX)}$, combined with the lift produced by the wing and tail, $F_{LZ}$, are less than the aircraft's weight, thus the aircraft is not in equilibrium. Hence, the stability region must lie above this constraint line.

## 4.3 Maximum Power
The power requirements on the engine exhibit a proportionality to the forward flight speed. As velocity increases, the oncoming blades experience higher velocity, leading to a quadratic increase in drag. While the propeller's thrust can be sustained, as this also scales quadratically, the power necessary to overcome drag escalates, eventually reaching a threshold where it cannot be maintained. Figure 2 depicts this phenomenon, through the representation of the maximum power line on the conversion corridor. This signifies the region by which the engine cannot sustain a sufficient level of thrust to satisfy Eq. (5).

## 4.4 Velocity Constraint
The velocity constraint describes the maximum velocity that the aircraft is designed to achieve. Due to the large blades, 'whirl flutter' can occur at high speeds [13], a phenomenon where the rotors induce resonant aeroelastic instabilities that can reduce efficiency and cause catastrophic failure of the aircraft. However, the exact nature of this phenomenon adds an additional layer of complexity to our aircraft, and thus is outside the scope of this paper, hence the velocity constraint was formulated from historic data.

## 4.5 Safety Margin
Incorporating a safety margin into our conversion corridor serves to alleviate the impact of potential errors in our formulation. While various mathematical methods have been employed to derive the data presented in figure 2, significant uncertainties persist, due to the utilisation of simplified models for thrust and lift. Introducing a safety factor into the formulation reduces the safe operating area, but ensures that through most of the transition, stability is feasible. However, as the final stability point should lie within this safety margin, uncertainty closer to the end of transition will persist. This is due to our requirement to control height, leading to a single trim position at the border of the wing stall constraint, hence the end stability point is located within the margin of safety.

## 4.6 Observations & Control
Given the derived safe operating area, we can select several points, within the region, as design points for our control simulation [14]. At each point, we will tune the PID gains to achieve the best result and then use gain scheduling to tie the design points together, as the aircraft transitions.

## 5. CONTROL IMPLEMENTATION
The following section describes the derivation of a control methodology aimed at regulating four key parameters: rotor angle, pitch angle, altitude, and velocity. By exerting control over these variables, our objective is to maintain the aircraft in straight and level flight throughout its transition, thereby fulfilling our overarching project aim of developing a robust control method for the transition mode.

The model is formulated under several key assumptions:

1. **Uniform Tilt Angle**: *The tilt angle of each rotor is assumed to be identical.*
2. **Uniform Thrust Output**: *Thrust is assumed not to be a function of tilt rate.*
3. **Non-Negative Thrust**: *The thrust generated by the propulsion system is constrained to be non-negative.*
4. **Symmetric Thrust Generation**: *The left and right-side engines are assumed to generate an equal amount of thrust.*

## 5.1 Rotor Angle Control
From the Conversion Corridor established in section 4, for varying forward flight speeds, we have defined a range of stable tilt angles. As such we will define a flight profile within this defined region of the conversion corridor and discretise it into several sections. We will then use this as the basis to define a desired tilt angle, where a simple closed loop proportional controller will be used to moderate the tilting angle toward its desired position.

## 5.2 Pitch Control
For the duration of transition, large variations in pitch will cause large instabilities due to its coupled effect on thrust vectoring, as shown in Eq. (2). To negate this effect, the initial pitch condition is set to 0 degrees, with an objective to stabilise any disturbance and ensure sufficient sensitivity throughout the transition. This will be implemented with a gain scheduled PID approach.

## 5.3 Altitude & Velocity Control

In helicopter mode, altitude-based control will be employed to stabilise the system, given its influence on the vertical dynamics. However, in fixed-wing mode, lift generated by the wing becomes the primary determinant of height, and this lift force is directly proportional to the aircraft's velocity. Thus, during fixed-wing mode, control based on forward velocity becomes more relevant. Therefore, the objective of this method is to transition from altitude control to forward velocity control during the transition mode, aiming to enhance performance in the control simulation.

## 5.4 Error System

To control each parameter, we will define an error term $e(t)$ as the difference between the desired state of the system and the actual measured state of the system, where error is a function of time. The error terms for altitude, velocity, and pitch $[e_Z(t), e_{\dot{X}}(t), e_\theta(t)]$ are shown in Eq. (6), where the desired altitude and pitch angle are defined as 0 and the desired velocity, $\dot{X}_d(t)$ is a function of time.

$$\begin{bmatrix} e_{\dot{X}}(t) \\ e_Z(t) \\ e_\theta(t) \end{bmatrix} = \begin{bmatrix} \dot{X}_d(t) \\ 0 \\ 0 \end{bmatrix} - \begin{bmatrix} \dot{X}(t) \\ Z(t) \\ \theta(t) \end{bmatrix} \quad (6)$$

The objective of this method is to make the error terms tend to zero, within an amount of time to maintain the aircraft stability.

## 5.5 Error Derivatives

The derivatives of Eq. (6) are taken to produce Eq. (7), where the right-hand side (RHS) of the equation defines the linear and rotational acceleration of the aircraft.

$$\begin{bmatrix} \dot{e}_{\dot{X}}(t) \\ \ddot{e}_Z(t) \\ \ddot{e}_\theta(t) \end{bmatrix} = - \begin{bmatrix} \ddot{X}(t) \\ \ddot{Z}(t) \\ \ddot{\theta}(t) \end{bmatrix} \quad (7)$$

## 5.6 System Dynamics

Modifying Eq. (7) by introducing the mass, M, and inertia, J, results in a force term equal to the forces and moments acting on the aircraft, presented in section 3.1, this can then be equated to the error derivatives, resulting in Eq. (8).

$$\begin{bmatrix} M\dot{e}_{\dot{X}}(t) \\ M\ddot{e}_Z(t) \\ J\ddot{e}_\theta(t) \end{bmatrix} = - \begin{bmatrix} M\ddot{X}(t) \\ M\ddot{Z}(t) \\ J\ddot{\theta}(t) \end{bmatrix} = \begin{bmatrix} F_X \\ F_Z \\ T_\theta \end{bmatrix} \quad (8)$$

Then finally, substituting the left-hand side of this equation into the aircraft mathematical model,(section 3.6) results in Eq. (9).

$$-\begin{bmatrix} M\dot{e}_{\dot{X}}(t) \\ M\ddot{e}_Z(t) \\ J\ddot{e}_\theta(t) \end{bmatrix} = \begin{bmatrix} F_{LX} \\ F_{LZ} \\ M_{L\theta} \end{bmatrix} + \begin{bmatrix} F_{TX} \\ F_{TZ} \\ M_{T\theta} \end{bmatrix} + \begin{bmatrix} 0 \\ 0 \\ M_{\tau\theta} \end{bmatrix}$$
$$+ \begin{bmatrix} 0 \\ W \\ 0 \end{bmatrix} + \begin{bmatrix} \delta_X \\ \delta_Z \\ \delta_\theta \end{bmatrix} \quad (9)$$

## 5.7 Controller Formulation

Whilst Eq. (9) describes a theoretical system where the error in: altitude, velocity and pitch, tend to zero, it has no method of imparting force onto the system nor any method of reducing error. Solving Eq. (9) for the controllable forces, defined in section 3.3 ($F_1$ and $F_2$), we can solve for the required thrust at any time. Then substituting an autonomously stable system, into the error derivatives, produces a complete control method. However, looking at Eq. (2), it is evident that $F_1$ and $F_2$ cannot be combined into a single variable that satisfies all three control equations. For this reason, we must define temporary control parameters for each control equation, then formulate a method to convert these temporary control parameters, into real values of thrust output, that satisfy all three equations.

## 5.8 Autonomously Stable System

We have formulated a controller by which the control forces $F_1$ and $F_2$, stabilise the system, though the error variables are still unknown. Thus, we have selected the autonomously stable differential equation, defined in Eq. (10). The benefit of using this system is that it is a differential equation whose solution is a decay function determined by the appropriate selection of the three gains: $K_P$, $K_I$ and $K_D$. Hence selecting this system autonomously causes the error to decay over time, thus, system stability can be achieved by the appropriate selection of each of these constants for each control variable.

$$\ddot{e}(t) = -K_P e(t) - K_I \int e(t)\, dt - K_D \dot{e}(t) \quad (10)$$

The velocity error is in a different form than that described in Eq. (10). So, an alternative form of this equation is used in the control of the velocity.

## 5.9 Temporary Controller

A solution for the force outputs of Eq. (9) are presented in Eq. (11) & (12) and expressed as a linear combination of temporary control parameters. These temporary control parameters are simply the grouping of terms from Eq. (9) and represent the control requirement from the altitude and position dynamics, $C_{z+\dot{x}}$ as well as the control requirement from the pitch dynamics, $C_\theta$.

$$F_1 = \frac{C_{z+\dot{x}}}{g(\tau)} + \frac{C_\theta}{g(\tau) + h(\tau)} \quad (11)$$

$$F_2 = \frac{C_{z+\dot{x}}}{h(\tau)} - \frac{C_\theta}{g(\tau) + h(\tau)} \quad (12)$$

Here, the systems equilibrium positions are satisfied, as when the control requirement of pitch is zero, the thrust outputs relationships are determined based on each engine's respective lever arm.

### 5.10 Parameter Selection
When selecting: $K_P, K_I$ and $K_D$, for each controller, we must obey the Routh-Hurwitz stability criterion [15], for which a routh table was generated. Notable observations of the stability criterion led to the conclusion, that the variations in the pitch and rotor angle exert an influence on the bounds of stability. This phenomenon, implies that the responsiveness of the controller is subject to variation, depending on the prevailing system conditions.

**Table 2 Control Gains for Helicopter and Fixed-Wing Modes**

|  | Helicopter Mode | Fixed-Wing Mode |
|---|---|---|
| $k_{pz}$ | 0.07 | 0.31 |
| $k_{iz}$ | 0.003 | 0.052 |
| $k_{dz}$ | 0.7 | 0.72 |
| $k_{p\dot{x}}$ | N/A | 1.1 |
| $k_{i\dot{x}}$ | N/A | 0.5 |
| $k_{d\dot{x}}$ | N/A | 0.7 |
| $k_{p\theta}$ | 0.29 | 0.6 |
| $k_{i\theta}$ | 0.0018 | 0.08 |
| $k_{d\theta}$ | 0.5 | 0.6 |

## 6. SIMULATION
The control simulation was developed in MATLAB, SIMULINK Version 2021B.

### 6.1 Helicopter & Fixed Wing
A simulation was performed for the fixed-wing and helicopter operating modes. A step input of 100 metres in height and 6 degrees in pitch was used as a baseline to assess the performance of these two modes.

### 6.2 Transition Mode
The main simulation will focus on the demonstration of the derived transition mode control system where Figure 3 shows how the derivations within the previous sections have been used to build the control simulation.

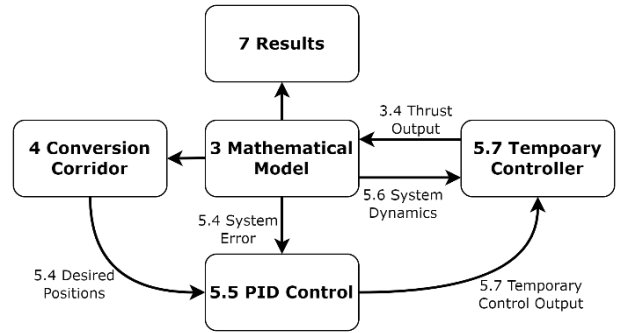

**Figure 3 Control Simulation Design**

## 7. RESULTS & DISCUSSION
Whilst the aim of the project lies in the stability of the transition mode, it's crucial to recognize that the chosen aircraft configuration must perform adequately in all three of its designed operating modes. Since the scope of this project is assessing the transition mode for a developed aircraft configuration, invalidating the configuration with instability in any of its operating modes would not be conducive to a robust control design.

### 7.1 Altitude Control
The altitude response for the step input, see section 6.1, are shown in Figure 4, where we see both modes capable of reaching stability at its desired position.

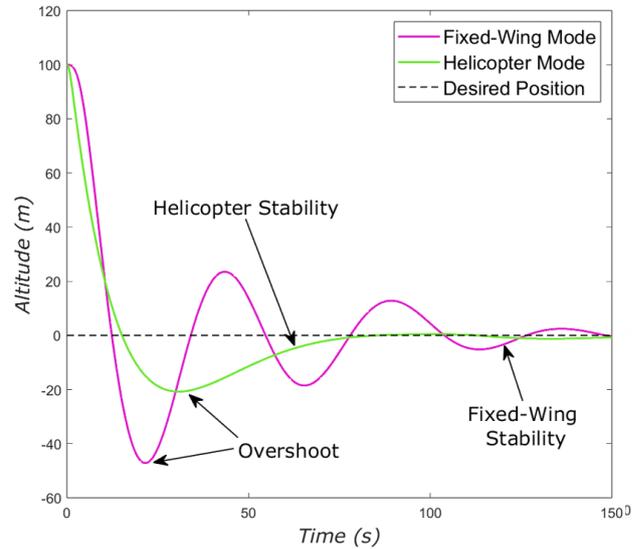

**Figure 4 Altitude Response to Step Input**

The more violent, and oscillatory behaviours of the fixed-wing mode, is partly due to the coupled effects of altitude and velocity control. The performance in helicopter mode is shown to be better, with lower overshoot and reduced oscillations, though the time to reach stability, in both cases, is significantly large. Whilst this behaviour was observed from the respective control gains, shown in Table 2, greater performance may be achievable, though unnecessary for our application. Notable is that the method to select appropriate control gains was sufficient to attain a reasonable stability point but was not capable of determining the most optimal control.

## 7.2 Velocity Control

The velocity fluctuations, $V$, were normalised by their respective trim velocities, $V_{Trim}$, as per Eq. (13). Results are shown in Figure 5, where we see both control methods capable of directing the aircraft to respective trim velocities.

$$V_{Normalised} = \frac{V}{V_{Trim}} \qquad (13)$$

For helicopter mode, velocity stabilisation occurs, despite only altitude control methods, though the trim velocity changes post step input. Notable here, is unlike fixed-wing mode, helicopter mode has a range of stable trim velocities. As no velocity control is implemented the control loop has no method to return the aircraft to its original velocity, however, this could be achieved by setting a desired trim pitch angle instead.

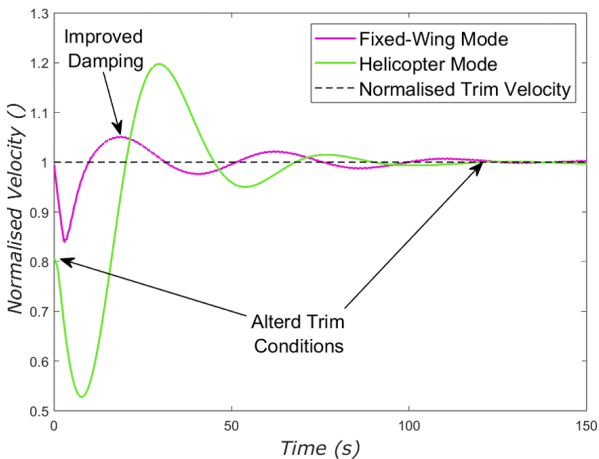

**Figure 5 Velocity Response to Step Input**

The introduction of a velocity controller for fixed-wing mode, is shown to dampen oscillations more significantly. Conversely, the velocity for helicopter mode has large fluctuations, though the fast damping of velocity is less important due to its minimal effect on system stability. With further iterations, the performance could be improved, however, the aim of a successful transition mode control only requires stable control for the other two modes. Thus, the stability shown in the data concludes that the controllers are adequate for our purposes.

## 7.3 Transition Mode Control

The output, of the transition mode control simulation, is illustrated in Figures 6, where various system states are shown. The simulation commences from helicopter mode, where the helicopter control parameters are utilized to bring the system to a stable trim position. Once the aircraft reaches this stable trim position, it is commanded to undergo transition, with rotor angle control activated. As the aircraft achieves a tilt angle of 85 degrees, the control switches from static helicopter control to gain-scheduled transition mode control, with static fixed-wing control being reinstated, only after the tilt angle drops below 5 degrees.

## 7.4 Altitude & Pitch Response

The results in Figure 6 (a & b) show the variations in altitude and pitch. Whilst the pitch angle is not zero, the observed variations are significantly low, (<0.1 degrees) that we would consider the aircraft stable. A notable increase occurs post transitioning, however the magnitude is small, further decaying over time, hence the variation does not pose a significant impact on our models' stability. The altitude variations are also small (<10m), but non-trivial, as it affects our assumption of zero flight path angle. It is seen that the fixed-wing control can stabilise the height, though further work is needed to characterize the effect of small flight path angle variations.

## 7.5 Longitudinal Velocity Response

Velocity increases rapidly over the transition period, where we see an almost linear increase. Like the altitude, the fixed-wing control stabilises the velocity post transitioning, however, there is a small discontinuity, caused by switching of control methods.

## 7.6 Tilt Angle Fluctuation

The tilt angle desired position, is governed by the conversion corridor set out in section 5. The maximum thrust is below its limits and the altitude variation is small which shows that our tilt angle control method, can keep the aircraft within stable bounds. Transitioning was achieved in under 14 seconds, notably similar to the bell XV-15, (12.5) [1], and whilst this value could be further decreased, this was not feasible due to the maximum thrust limits imposed by both the model, and the conversion corridor.

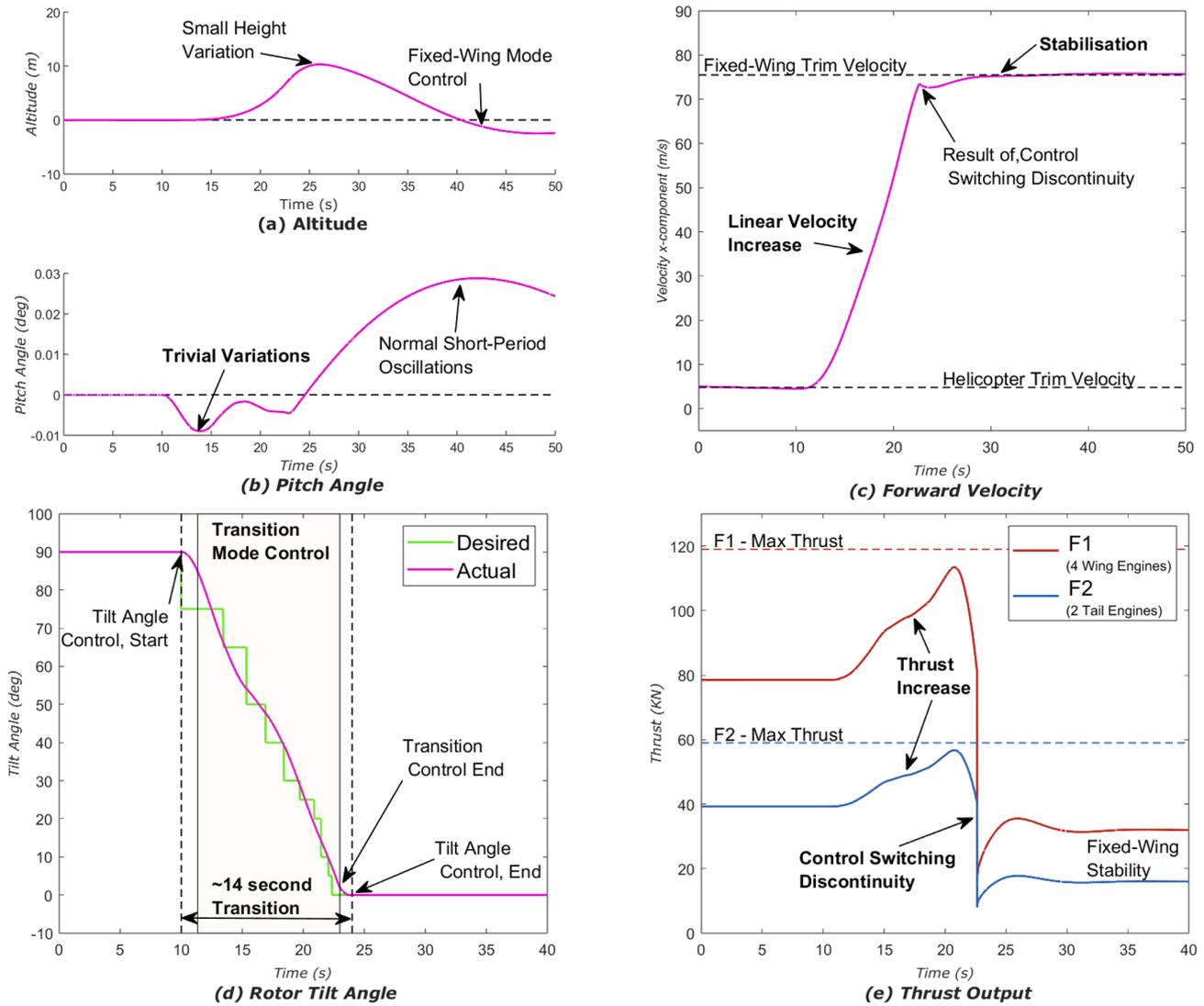

**Figure 6 Transition Mode, Control Simulation Performance Data**

### 7.7 Thrust Output Implications
For a 14 second tilt, the peak thrust observed, fell below the maximum thrust limits placed on each engine group. The thrust increases steadily, until a peak, at which point the lift force generated by the wings starts to overtake the thrust output. This is followed by a sharp decline, as the tilt angle approaches zero, directing more thrust forward, as well as the reduced thrust required by fixed-wing mode. However, a discontinuity occurs in this region, being a combination of rapidly changing thrust requirements coupled with switching control modes. More refinement in the switching of control methods here, would improve the simulation, though despite this, the system reaches a stable fixed-wing operating state. This is relevant, as the ability of our aircraft to achieve such violent thrust levels is an unknown. Notable here, for previous iterations with lower tilt rates, the maximum observed thrust was reduced. This observation is important, as by defining an upper thrust limit, we could effectively identify the maximum tilt rate. Though outside the scope of this paper, formulations of tiltrotors with defined transitioning requirements would benefit from this information.

### 7.8 Overall Stability
The results presented are important for our aims, as a holistic view of performance is required to assess the robustness of control. The controllable elements of thrust output and rotor angle are shown to produce minimal variation in the altitude and pitch, while also rapidly increasing and stabilising velocity. Thus, we can conclude our control implementation has been successful in completing a transition, however this is only valid for the given assumptions.

### 7.9 Simulation Improvements
Though the model has demonstrated its ability to perform transition, it has done this under idealized conditions. To improve the robustness of the control, further simulations should be able to characterise the effect of external disturbance to see what effect it has on performance, and at what point stability can no longer be achieved.

## 8. FURTHER WORK
### 8.1 Relaxation of Assumptions
In our mathematical model, we have made some assumptions. Whilst it is true that we have achieved stable control, this method and simulation is only valid for the set of assumptions outlined throughout this paper. We have evaluated that flightpath angle is non-zero, hence, the mathematical model would need to be re-evaluated to determine the effect of small variations.

### 8.2 Simulation Refinement
As suggested, the control switching discontinuity observed, could benefit from more localised control. The region could be further discretized, with more scheduled gains, however the increased complexity from this would necessitate a more refined method of gain scheduling.

### 8.3 Thrust Transition Rate
Whilst the performance of the aircraft has been captured in this paper, it would be beneficial to analyse the relationship between transitioning speed and thrust output. This would be a beneficial metric as to compare to other similar control methods, and would establish the limitations of our model.

## 9. CONCLUSION
In summary, this paper introduces a novel tiltrotor aircraft characterized by its large-size and configuration featuring six distributed engines. For this aircraft, a mathematical model has been produced, and the bounds of stability have been determined via the construction of a conversion corridor. A control method was then implemented to stabilise the aircraft during helicopter, fixed-wing, and transition mode, and through simulation, has demonstrated its ability to stabilise the aircraft. Although the altitude and pitch show minor variation, the overall aircraft stability is withheld over time. thus, for our given assumptions, we can conclude that we have advanced toward our aim of producing a robust control method. However, significant caveats to the validity of our controller remain, though with further refinements to the control, and analysis of relaxed parameters, the validity could be significantly improved.